# Enhanced clustering tendency of Cu-impurities with a number of oxygen vacancies in heavy carbon-loaded $TiO_2$ - the bulk and surface morphologies

D.A. Zatsepin[1,2], D.W. Boukhvalov[3,4], E.Z. Kurmaev[1,2], A.F. Zatsepin[2], S. S. Kim[5], N.V. Gavrilov[6], I.S. Zhidkov[2]

[1] M.N. Miheev Institute of Metal Physics of Ural Branch of Russian Academy of Sciences, 620990 Yekaterinburg, Russia
[2] Institute of Physics and Technology, Ural Federal University, 620002 Yekaterinburg, Russia
[3] Department of Chemistry, Hanyang University, 17 Haengdang-dong, Seongdong-gu, Seoul 04763, Korea
[4] Theoretical Physics and Applied Mathematics Department, Ural Federal University, Mira Street 19, 620002 Yekaterinburg, Russia
[5] School of Materials Science and Engineering, Inha University, Incheon 402-751, Republic of Korea
[6] Institute of Electrophysics, Russian Academy of Sciences, Ural Branch, 620990 Yekaterinburg, Russia

*The over threshold carbon-loadings (~ 50 at.%) of initial $TiO_2$-hosts and posterior Cu-sensitization (~ 7 at.%) was made using pulsed ion-implantation technique in sequential mode with 1 hour vacuum-idle cycle between sequential stages of embedding. The final $C_x$-$TiO_2$:Cu samples were qualified using XPS wide-scan elemental analysis, core-levels and valence band mappings. The results obtained were discussed on the theoretic background employing DFT-calculations. The combined XPS-and-DFT analysis allows to establish and prove the final formula of the synthesized samples as $C_x$-$TiO_2$:[$Cu^+$][$Cu^{2+}$] for the bulk and $C_x$-$TiO_2$:[$Cu^+$][$Cu^0$] for thin-films. It was demonstrated the in the mode of heavy carbon-loadings the remaining majority of neutral C–C bonds ($sp^3$-type) is dominating and only a lack of embedded carbon is fabricating the O–C=O clusters. No valence base-band width altering was established after sequential carbon-copper modification of the atomic structure of initial $TiO_2$-hosts except the dominating majority of Cu 3s states after Cu-sensitization. The crucial role of neutral carbon low-dimensional impurities as the precursors for the new phases growth was shown for Cu-sensitized $C_x$-$TiO_2$ intermediate-state hosts.*

# 1. Introduction

Transition metal (TM) oxide semiconductors with a relatively narrow band-gap (2.7 eV – 3.2 eV) are considered as an ideal host-materials for the area of photovoltaic applications because of their good chemical stability, non-toxicity, low-cost and well-developed conventional synthesis methods. Among TM-oxides the titanium dioxide, having several polymorphs, seems to be potentially promising as a base for highly efficient photocatalyst in water splitting [1-2], anti-fouling [1], the major component for the photostimulated degradation of pollutants [3-5], TM-sensitization of solar-cell production [6-7]. The TM-sensitization of $TiO_2$-host is considered to be highly effective for the photoactivity improvement of $TiO_2$ by means of decreasing the charge carriers recombination rates and, hence, the enhancement of photocatalytic features. Despite the reported relatively wide commercial applicative area for $TiO_2$, there are still several scientific and technological challenges which are impeding the expansion and improvement of $TiO_2$-host functionality. Recently it was established, that an optimization attempts of $TiO_2$ functional properties with the use of conventional techniques lead in some cases to even a bit wider band-gap value than that for untreated host-matrix, thus indicating the incompatible with photo-energetics area the final electronic structure of the functionalized host [8-9]. For instance, Choudhury et. al. was reporting about the strong re-arrangement of initial $TiO_2$ atomic and electronic structure after TM-sensitization, linking this phenomenum with Jahn-Teller distortions of $TiO_6$-octahedra, caused by appeared oxygen vacancies (both point and hollow-types), strong *d-d* excitations and conversion from the octahedral symmetry to tetragonal one [10]. As a separate case of matter the rutile-to-anatase temperature stimulated junctions were reported and discussed as well. So it becomes clear that the functional properties of TM-sensitized $TiO_2$-host might be essentially affected by a large number of factors which are including the particle size (microcrystallinity type), the active surface area of the fabricated particles, the morphology of the material, etc. – viz. *in-situ* synthesis method and conditions. Wherein one from the major challenging questions is the point how to modify the energy band-gap in order to coincide well with the requirements of an "ideal" photocatalyst [11-13] and, at the same time, not to distort the residual functionality of a modified host by suppressing the possible recombination of photoinduced electrons and holes under visible light excitation.

An applicative importance of titanium dioxide as a host-material for the structure sensitization in terms of photoactivity is following from the fact that this material is allowing to employ the relatively wide range of processing techniques with promising final results. For instance, an electrochemical synthesis (an anodizing of titanium in different solutions and doping with transition metals [14]) seems to be promising because of the yielded final functionalized $TiO_2$-structure with enhanced photocurrent [15-16], deceleration of the photocatalytic degradation side-effects [17] and, thus, elongation of performance cycle-time, in some cases of corrosion protection [18] of an active photocatalyst electrode zones. With that, this materials science area seems to us as still challenging because of the need to apply actually specific solvents [16,19]; appearance of oxygen sublattice defects in the sensitized $TiO_2$ [19]; the unwanted variety of the charge states of an employed for the sensitization transition-metal dopant [20].

As a relatively novel and effective approach for $TiO_2$ sensitization with TM's, an additional doping with carbon might be considered [21-27]. The most remarkable conclusions, following form these cited papers allow to point the reduction of $TiO_2$ microparticles dimension with simultaneous enlarging of their photoactive surface (the so-called synergistic effect, see i.e. Refs. [28-30]), higher absorptive ability of the incident light and the re-distribution of photo-generated charge-carriers with the overall improvement of energy conversion efficiency. Unfortunately an excess of embedded carbon via high-concentration mode of carbon-loadings of host-matrix might introduce concrete and purely negative side-effect: the increasing the overall electric resistivity of a carbon-doped photoactive material with the onward degradation of photoinduced current up to negligibly small value [31]. In order to avoid the overall electric resistivity increasing of C–$TiO_2$ host, the co-doping with a good TM-conductor might be considered, i.e. metal copper, but this will increase of amorphous grain boundaries with all the consequences (at least while employing conventional doping approaches) [10]. Thus the situation with carbon-copper co-doping of TiO$_2$-host is not so straightforward and the combined theoretical and experimental study of this question is needed for the deeper clearance.

In the current paper we will present the examined and established influence of heavy carbon-loadings (~ 50 at.%) and posterior Cu-sensitization (~ 7 at.%) on the electronic structure and defects of initial *bulk* and *thin-film* (surface) morphologies of TiO$_2$-hosts. As previously, the combined X-ray Photoelectron Spectroscopy (core-levels and Valence Band) and DFT-based modeling approach will be employed. The key-point attention will be paid to the clustering tendency of embedded carbon and copper implants as well as to the oxygen sublattice re-arrangements. This pending case of TiO$_2$ carbon-loadings, being the characteristic feature of applied ion-implantation techniques, essentially exceeds the standard chemical solubility limits of conventional carbon-doping modes (over the threshold saturation) for titanium dioxide and, thus, might be of special interest for different technological applications.

## 2. Samples preparation and experimental details

For the subsequent ion-implantation treatment of TiO$_2$ as a host-matrix, the *bulk ceramics* and *thin-film* morphologies of titanium dioxide were employed as previously. Recall, that using sample the preparation methodology, XRD, and preliminary XPS characterizations, reported in Refs. [32-34], the single-phase phase rutile with lattice parameters of $a = 4.592$Å and $c = 2.960$ Å and average crystallite size of about 200 nm (*bulk ceramics* or from now simply *bulk*) and single-phase anatase *thin-films* [32] with lattice parameters of $a = 3.785$Å and $c = 9.487$ Å were obtained.

The exordial brief analysis of third-party research data allows to derive that the carbon-copper co-implantation treatment seems to be the logically justified choice for employed TiO$_2$-hosts. Since the simultaneous pulsed co-implantation with two types of ions adducts to the essential conversion of an oxygen-ligand and central atom sub-lattices (see i.e. Ref. [35]), thus one cannot assume that this mode of implantation will be suitable for the electronic structure soft modulation (ESM) [36] of our TiO$_2$-hosts [37]. So we will make an attempt to perform the sequential step-by-step pulsed ion-implantation firstly with carbon (so-called Cx→TiO$_2$ loadings [38]), but without hydrophobic treatment of the surface which usually causes the appearance of "alien" OH$^-$ groups and then with copper. The main idea of such an approach is to prevent or, at least, to reduce the initial host-structure total conversion for better compatibility with technological ESM-conditions.

The sequential carbon (first step of separate embedding) and then copper (second separate step) ion-implantations of TiO$_2$-hosts in a pulsed-repetitive mode per each step were made at a vacuum pressure of $3 \times 10^{-3}$ Pa. The MEVVA-type ion-source provided the identical embedding conditions both for C-ions and, after that, for Cu-ions: ion-energy of 30 keV, the pulsed-beam current density of 0.65 mA/cm$^2$, pulse repetition rate of 12 Hz, a pulse duration of 0.3 ms and flux of $1 \times 10^{17}$ cm$^{-2}$. Total time-exposure under appropriate ion-beam treatment was not less than 5 hours for primary carbon embedding and then 70 minutes for the onward copper embedding with 1 hour of an idle-cycle between carbon and copper loadings. The average temperature of the host-samples under ion-embedding did not exceed 300°C. The earliest and detailed report about principles of the above briefly described technique and the current state of matter are presented in Refs. [39-40], respectively.

X-ray photoelectron qualification of the final electronic structure of sequentially carbon-copper ion-beams treated TiO$_2$-hosts was performed using a PHI XPS Versaprobe 5000™ spectrometer (ULVAC–Physical Electronics, USA) based on a classic X-ray optic scheme with a quartz monochromator and hemispherical energy analyzer working in the range of binding energies from 0 to 1500 eV with an employment of Al $K\alpha$ radiation (1486.6 eV, $\Delta E_{resolution} \leq 0.5$ eV, 100 μm X-ray probe dia, X-ray power loading of the samples less than 70 Watts). More detailed description of the used XPS recording parameters might be found in our previous publications (see, i.e. Refs. [33-35]). For a complete description of PHI XPS Versaprobe 5000™ system, please, see the web-site of spectrometer manufacturer. According to the accepted ASTM Standards of XPS measurements and charge-referencing [41] the complete electronic structure mapping – fast wide-scan (survey), core-levels and valence bands – had been made. The interpretation of the XPS data obtained was performed with the crossed-check referencing of internationally accepted XPS Databases [42-43] and Ref. [44]. No argon sputtering of the samples under study was made to prevent an inadvertently sputter-induced fabrication of metal carbides and carbonates which might occur under this kind of treatment in carbon-loaded metal-containing oxides, nitrides, borides, etc. [43]. As a separate and additional testimony that no TiC or/and other titanium-carbon derivative compounds are present in our host-samples after carbon-ion loadings an exactly high binding energy difference between the Ti 2p XPS for $TiO_2$ and TiC can be considered – about 5.3 eV [45]. This noticeable difference will be clearly visible seen even in the XPS wide-scan spectra but it is absent, and all Ti 2p for our implanted hosts and reference $TiO_2$ have the same BE-positions within the spectral resolution of recorded XPS wide-scans (see Fig. 1).

The comparison of XPS fast wide-scan qualification (XPS chemical elements analysis) for sequentially carbon-copper ion-beams modulated $TiO_2$-hosts in the *bulk,* and *thin-film* morphologies are shown in Fig. 1. One can see that the XPS wide-scans of the samples under study are characterized by the appearance of additional specific XPS peaks, belonging to the appropriate copper electronic states. Moreover, much higher C 1s signal is manifesting in compliance with $TiO_2$ XPS external standard. The detected spectral hangs are not abrupting because exactly these elements had been sequentially implanted into TiO$_2$-hosts. Leastways, around three times higher intensity of C 1s in the wide-scans of Cx-TiO$_2$:Cu samples in both employed morphologies versus reference TiO$_2$ might be recognized as a sequence of performed carbon loading via first ion-implantation step. Nevertheless, there is a chance to speculate reasonably about vacuum hydrocarbons contamination of our samples which usually results in distortions of XPS C 1s simple-shape symmetry as well as emerging additional well-known XPS sub-bands in the range of C 1s binding energies. To clarify this case of matter the high-resolution C 1s core-level XPS analysis was made (see Fig. 2).

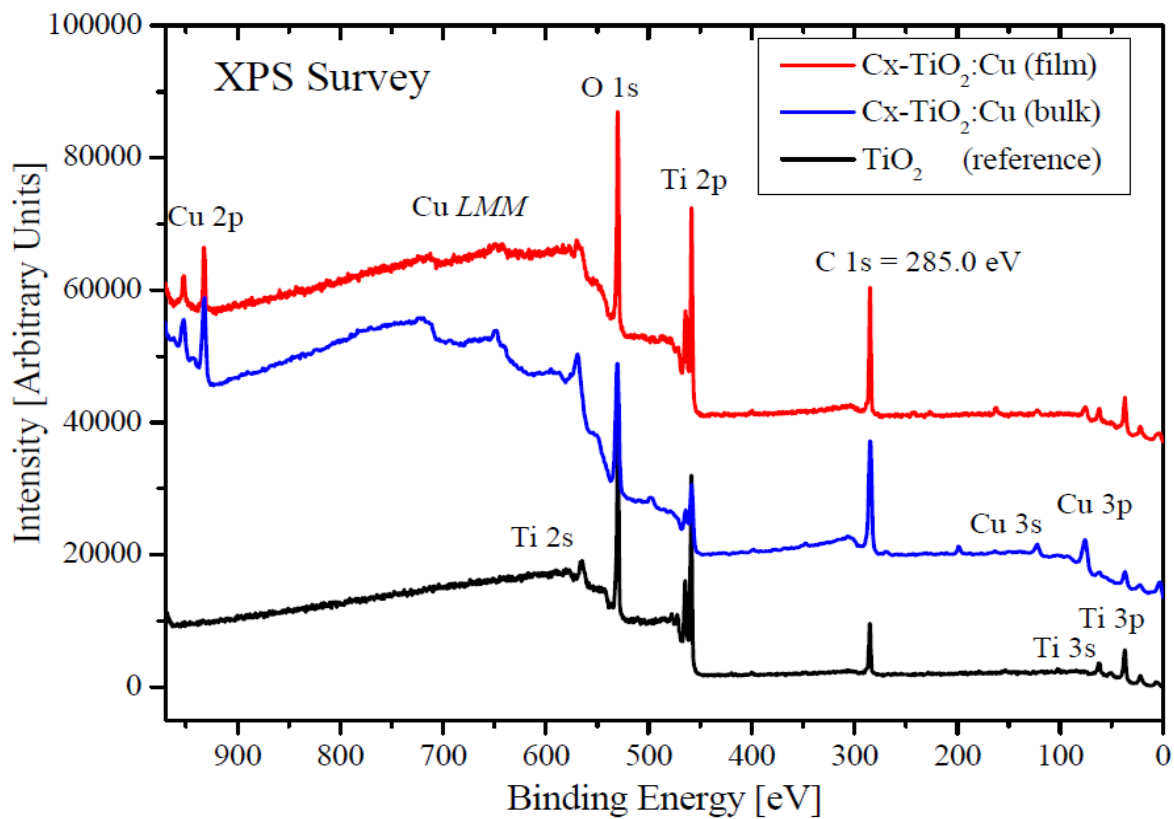

**Figure 1.** XPS fast wide-scan (survey) elements analysis of the carbon-copper modulated $TiO_2$-hosts in the *bulk* and *thin-film* morphologies with referencing to $TiO_2$ XPS external standard.

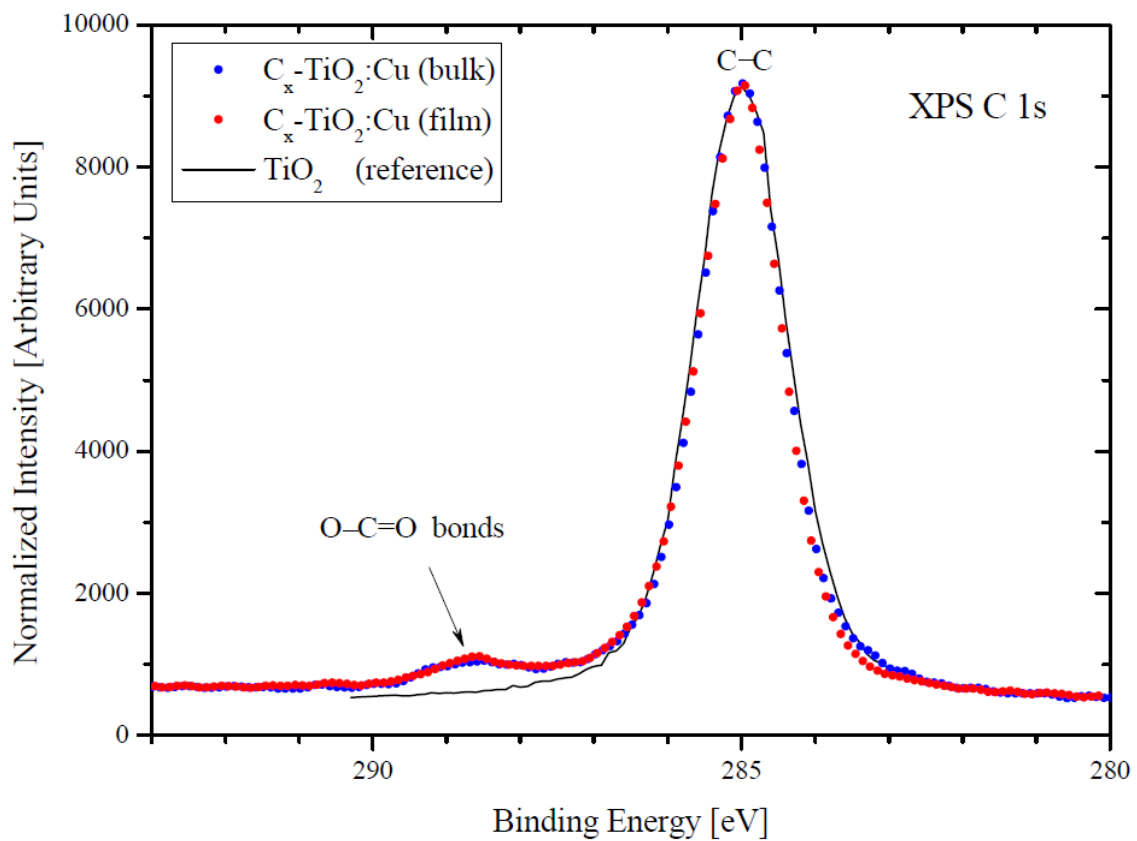

**Figure 2.** XPS C 1s core-level spectra of the carbon-copper modulated TiO$_2$-hosts in the *bulk* and *thin-film* morphologies with the referencing to TiO$_2$ XPS external standard.

From Figure 2 one can clearly see, that C 1s core-levels in the electronic structure of Cx-TiO$_2$:Cu samples and TiO$_2$ reference have the identical in shape-symmetry XPS main-lines located at 285 eV which represent the C–C bonding. With that, there are no any additional spectral contributions in the vicinity of 286 eV binding energy value, thus no chance for the statements about C–O–C type of contaminations or/and imperfections, because they are manifesting themselves, in general, as a visually asymmetrical shoulder exactly in this BE-region for the wide set of carbon-oxygen containing materials [42-43]. At the same time, the 288.7 eV XPS sub-band is present only in the carbon-copper modulated $C_x$-$TiO_2$:Cu samples with no even a sign of it in the C 1s core-level of $TiO_2$-reference (Fig. 2). Thus the character of 288.7 eV XPS sub-band manifestation, in our humble opinion, might be linked with the aftermath of carbon loadings via ion-implantation and not with vacuum hydrocarbons contamination. According to XPS Databases, this sub-band is arising due to the irregular O–C=O fragments [43] thus supporting our derivation about C-implanting reason for it. Recall that the 288.7 eV XPS sub-band is absent at all in C 1s spectrum of $TiO_2$ reference, which was not carbon-implanted, but only surface-marked with adventitious carbon for the proper charge referencing (see Fig. 2). From above one might conclude rather about an incomplete interaction of implantation- embedded carbon with titanium-oxygen lattice than about vacuum hydrocarbon contaminators, as it was reported in Ref. [42]. Again, no any XPS signals were found in the range of 282 eV in C 1s core-levels of the samples under study, so this is one more evidence that no TiC and/or other titanium-carbon derivatives were fabricated unexpectedly in our samples because of the improper and unsuitable technological treatments [45].

Finally, the performed XPS qualification allows to state that no alien XPS components are present in XPS wide-scans and C 1s spectra, thus proving the absence of contaminators in $C_x$-$TiO_2$:Cu final samples. The averagely estimated concentrations of injected via implantation carbon and copper are shown in Table I.

**Table I**. Estimated concentrations of components in sequentially carbon-copper implanted $TiO_2$-hosts.

| Sample (morphology) | Concentration (at. %) | | | |
| --- | --- | --- | --- | --- |
|  | C | O | Cu | Ti |
| $C_x$-$TiO_2$:Cu (bulk) | 45.6 | 32.8 | 6.9 | 14.7 |
| $C_x$-$TiO_2$:Cu (film) | 43.3 | 33.8 | 6.7 | 16.2 |

## 3. Computational method and models

Density-functional theory (DFT) calculations were performed using the SIESTA pseudopotential code [46] as had been used successfully for related studies of impurities in the *bulk* and *thin-film* morphologies of TiO$_2$ [40,41]. All calculations had been made employing the Perdew-Burke-Ernzerhof variant of the generalized gradient approximation (GGA-PBE) [47] for the exchange-correlation potential accounting the Dipole Correction [48] as well. After that the calculated atomic positions were completely optimized. The ground electronic state was consistently found during optimization using norm-conserving pseudopotentials [49] for the cores and a double-ξ plus polarization basis of localized orbitals for C, Cu, Ti, and O. The forces and total energies were optimized with an accuracy of 0.04 eV/Å and 1.0 meV, respectively. All calculations were carried out with an energy mesh cut-off of 300 Ry and a *k*-point mesh of 6×6×4 in the Monkhorst-Pack scheme for the *bulk* and 6×6×2 for the "surface" (*thin-film*) [50]. In order to plot the final Densities of States (DOSes), the *k*-point mesh was increased up to 8×8×6 and 8×8×4, respectively.

The employed for the model calculations supercells were including 96 atoms both for the *bulk* rutile-TiO$_2$ and for the *thin-film* (surface) anatase-TiO$_2$, but in the latter case the supercell was used as a slab because it is the most physically feasible model of (001) TiO$_2$ surface [33,34]. Additionally, for the *bulk* morphology used in experiment both rutile and anatase polymorphic structures were theoretically tested and carefully examined, whereas for *thin-film* (surface) the anatase polymorphic structure was selected solely (see the reason in experimental part above).

The calculations of the formation energies ($E_{form}$) were based on the following formula:

$$E_{form} = [E_{total} - (E_{matrix} - nE_{removed} + mE_{added})]/m \qquad (1)$$

where $E_{total}$ – is the total energy of the system with the current configuration of defects; $E_{matrix}$ – denotes the total energy of the system before substituion of host-atoms with impurities, or the fabrication of vacancy or/and interstitial defects (the conversion mode of structural configuration), $E_{removed}$ and $E_{added}$ – are the energies of removed and added atoms which were calculated for the bulk

(graphite state for carbon) and gaseous phases (triplet state for oxygen) in the ground configuration; n and m – are the numbers of removed and added atoms, respectively. We have to note, that for the structures with multiple charge-state defects (i.e. embedded Cu-impurities and simultaneous presence of oxygen vacancies), $E_{\text{matrix}}$ already denotes the energy of TiO$_2$ supercell with oxygen vacancies.

Based on our previous theoretical experience (see Refs. [33,34]), we decided to consider as possible copper impurities the following types of defects: substitutional (S), interstitial (I) and their combination of (xS+I) type. The case of multiple defects combination was analysed and examined separately, employing different distances between various defects – i.e. among the oxygen vacancy, carbon and copper impurities). The calculations of the formation energies were performed only for the configurations with the lowest total energies of the systems.

## 4. Results and Discussion

From above performed XPS qualification of C 1s core-levels in the electronic structure of Cx-TiO$_2$:Cu samples in both studied morphologies becomes clear, that the majority contribution of carbon into our hosts turns out in the sp$^3$-bonded form [50]. This means that there were weak and insufficient interactions of carbon with titanium-oxygen lattice which results, with the high probability, only in the fabrication of O–C=O clusters, and in general, an embedded carbon remains in the form of neutral C–C bonds. No any spectral signatures of carbon-titanium interactions were found (see discussion above). The XPS asymmetrical and weak contribution of O–C=O XPS bands into C 1s core-levels at 288.7 eV (see Fig. 2) allows to suppose about the minority holdings among oxygen-carbon with sp$^2$-character of the latter, because all sp$^2$-bonded carbon compounds have definitely asymmetrical XPS line-shapes of C 1s core-levels in compliance with high-symmetrical XPS line-shapes for sp$^3$-bonding [43]. Thus from established dominating neutrality of embedded carbon logically follows, that the majority of the "host oxygen – implanted element" interactions might be accounted for the sequentially embedded copper, so we will analyze now the XPS Cu 2p core-levels of our samples which are shown in Fig. 3.

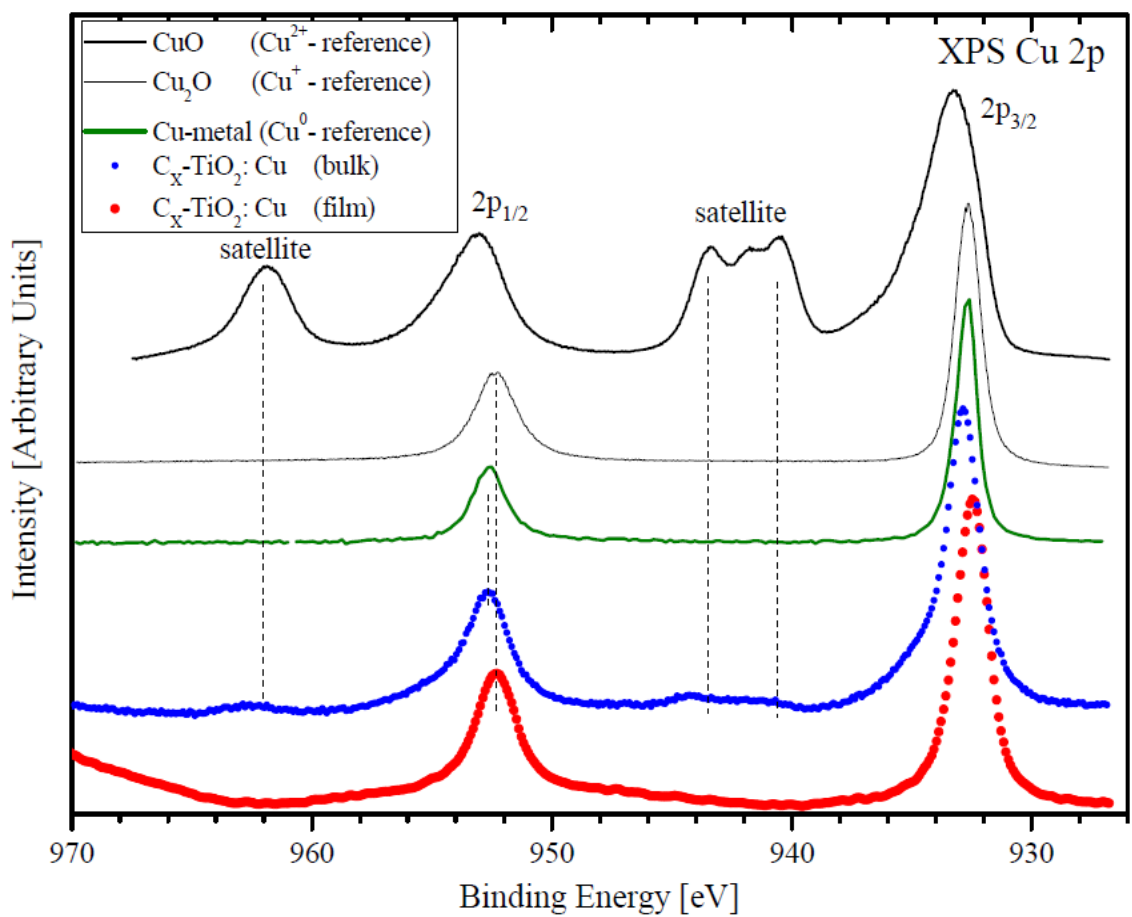

**Figure 3.** XPS Cu 2p core-level spectra of the carbon-copper modulated TiO$_2$-hosts in the *bulk* and *thin-film* morphologies with the appropriate XPS external standards: CuO, Cu$_2$O and Cu-metal.

One can see from Figure 2, that the main difference in XPS Cu 2p$_{3/2–1/2}$ for Cx-TiO$_2$:Cu in both employed morphologies is manifested in the clearly visible BE-shift of 2p$_{3/2–1/2}$ components relatively to each other and weak-intensity broad spectral features located in the 941-943 eV range and at 962 eV, respectively. As another visually observable feature of the *bulk* Cx-TiO$_2$:Cu XPS Cu 2p$_{3/2–1/2}$ might be considered the strongest asymmetry which is causing the shoulder at 934.7 eV, if comparing with that for *thin-film* (see Fig. 3, blue and red spectra). It seems that these mentioned above well-recognizable XPS spectral features are specific to establish the valence states of embedded copper ions into Cx-TiO$_2$ matrices. Formally we are dealing now exactly with Cx-TiO$_2$ hosts because the implantation was sequential and primarily the carbon was injected into initial TiO$_2$-hosts. On the strength of main-lines BE-positions in the XPS Cu 2p of our samples (Cu 2p$_{3/2}$ = 932.9 eV for the *bulk* and Cu 2p$_{3/2}$ = 933.4 eV for the *thin-film*) there are no background to link our XPS data with that for "copper-carbon" and "copper oxide-carbon" compounds (compare, i.e. in Cu$^{2+}$-carbonate di-hydroxide Cu 2p$_{3/2}$ = 934.7eV [43], in nanostructured copper phthalocyanine-sensitized multiwall carbon nanotube films Cu 2p$_{3/2}$ = 935.9 eV [51] and in composite copper oxide/chitosan/carbon fibers Cu 2p$_{3/2}$ = 935.7 eV [52]). Additionally, the shape of the characteristic "shake-up" satellites in these compounds is essentially dissimilar [43,52] with what was established in our XPS Cu 2p core-level analysis for both morphologies of Cx-TiO$_2$:Cu. Thus we have all the reasons to believe that no "copper-carbon" and "copper oxide-carbon" compounds were fabricated after sequential Cu-implantation of Cx-TiO$_2$ hosts on the second stage of ion-beam stimulated synthesis. Already well-known details about the XPS "shake-up" satellite and mixed valence states of copper in copper containing compounds are reported, for example, in Refs. [42-44, 53] or elsewhere. Finally, the high-oxidation and high-acidification abilities of copper (as well as of Pb, Sn, ..., etc. ) are excluding any residual doubts that copper-oxygen interaction is much easy than copper-carbon one, at least without any specially applied thermodynamic conditions which were absent in our synthesis.

From above discussion preamble the oxidation roots of Cu 2p core-levels shape-transformation in the electronic structure in Cx-TiO$_2$:Cu samples might be supposed as the basic background. Indeed, the XPS Cu 2p$_{3/2-1/2}$ for *thin-film* (Fig. 3, red spectrum) has the same BE-locations and line-shapes as Cu$_2$O and the absence of characteristic "shake-up" satellite, so formally the valence states of copper might be accounted as Cu$^+$. At the same time a little larger FWHM (more at ~ 0.3 eV) of the main XPS peaks in *thin-film* Cx-TiO$_2$:[Cu$^+$] than in Cu$_2$O possibly occurs due to partial contribution of metallic copper which has essentially neighbouring to Cu$_2$O binding energy values, making quite difficult to analyze the real state of matter in terms of copper charge states detection. This partial metallic Cu-loss effect has, at least, two valuable reasons in our humble opinion: (i) some oxygen-ligands from the titanium-oxygen lattice of the initial TiO$_2$-host had been previously "captured" by sp$^2$-bonded carbon;

(ii) an employed $1 \times 10^{17}$ cm$^{-2}$ ion-flux for copper embedding is the ride-line fork for the ion-beam stimulated processes of small metallic clusters growth and Oswald ripening (also known as particles coarsening) [54]. Both oxygen deficit after the first step of sequential carbon implantation and a relatively low atomic fraction of copper (see Table I) are reasonable enough to cause the appearance of metallic copper small clusters on the "growth stage" according to Ref. [54]. Thus they are contributing to the XPS Cu $2p_{3/2-1/2}$ core-level of *thin-film* as well, and we suppose the final formal formula as Cx-TiO$_2$:[Cu$^+$][Cu$^0$].

Another situation occurs with the *bulk* sample. Here one can see all the XPS signatures of the presence of Cu$^{2+}$ electronic states. They are: (i) visually noticeable asymmetry of Cu $2p_{3/2}$ main-line (usually absent in stoichiometric Cu$^+$-compounds); and (ii) the appearance of characteristic "shake-up" satellite (similar to CuO) in the same BE-region. Nevertheless, the main XPS lines for this sample are narrower than in ordinary (or natural) CuO, and the satellite is very weak (see Fig. 3). Moreover, the binding energies of the main lines are in better agreement with that for CuO$_2$ XPS external standard (compare thick black and blue spectra in Fig. 3), so this might be a sign of a mixed valence states of copper in the *bulk* morphology of Cx-TiO$_2$:Cu, namely Cx-TiO$_2$: [Cu$^+$][Cu$^{2+}$]. This is very close to that had been previously established by us and reported in Ref. [55]. The main dissimilarity of the XPS data reported in the current paper with that in Ref.[55] (in brief: spherical shock-waves loading by means of high-energy He$^+$ ions bombardment and shear under high-pressures of a native CuO as a model object) is that we got in the *bulk* Cx-TiO$_2$:Cu the very beginning of Cu$^{2+}$ valence states fabrication under low-energy sequential pulsed ion-beams treatment (ESM-mode) and the Cu$^+$-containing phase is dominating in the *bulk* sample. Finally, our XPS data on Cu $2p_{3/2-1/2}$ for Cx-TiO$_2$:Cu in both morphologies well agrees with that have been reported by Y.Xu *et. al.* [56] for solely Cu-doped TiO$_2$, thus proving our current and above reported findings that carbon is not affecting so much the charge states of copper in TiO$_2$ matrix and the valence states of Cu are determined exactly by the morphology of TiO$_2$ matrix and its oxygen ability to interact with embedded copper in our case of implantation. This conclusion does not seem precarious, because the same C-modulation conditions were applied both for TiO$_2$ *bulk* and *thin-film* hosts and, as a host reply result, the same reason for the manifestation of 288.7 eV XPS sub-band in these matrices occurs. With that, the morphology (*bulk* versus *thin-film*) and polymorphism (anatase versus rutile) of employed matrices are different because of the dissimilar titanium-oxygen unit-cell parameters in these polymorphs (slightly distorted cubic close-packing for O with Ti in a half of O$_h$ holes in anatase versus distorted hexagonal close-packing O with Ti in a half of O$_h$ holes in rutile [57]). The latter might be at least one of the possible valuable reasons for different type of Cu-embedding into the polymorphic hosts with the identical chemical-composition formula. This matrix-reply effect for the metal-implant with easily oxidized abilities was established by us as well for the Pb-modulated TiO$_2$-hosts being in the same morphologies [53] and does not contradict the current findings.

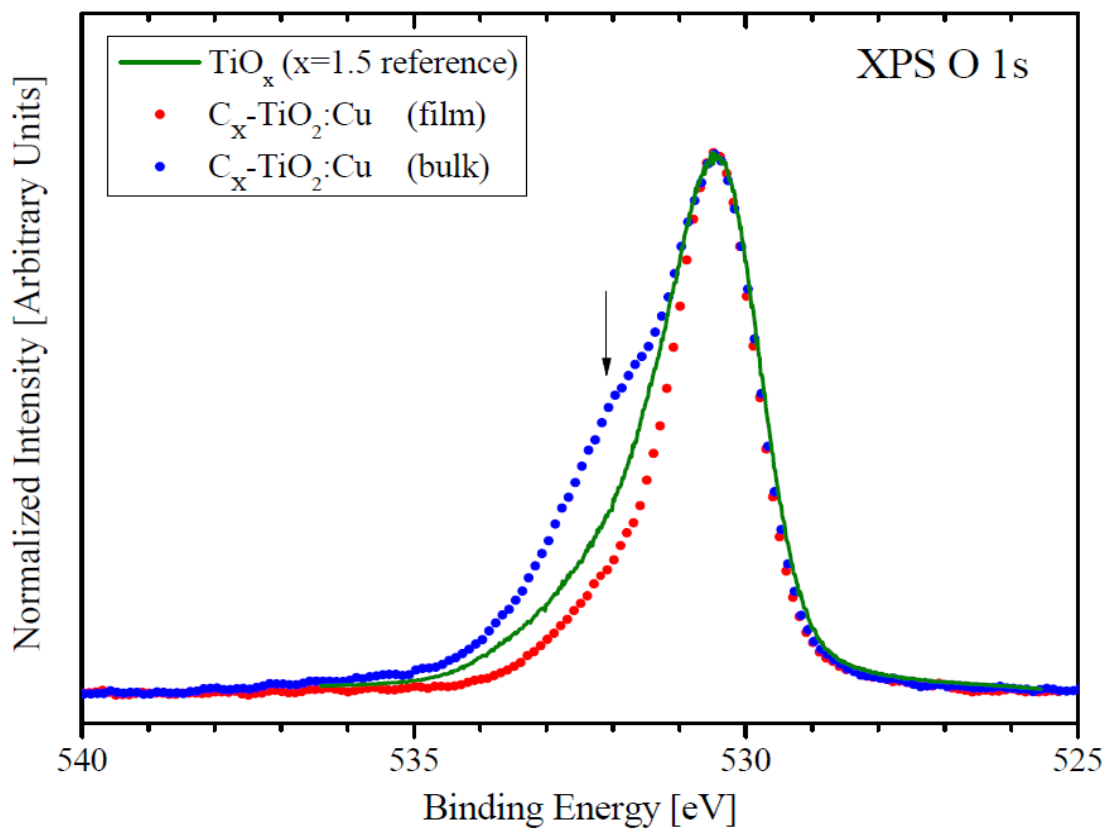

**Figure 4.** XPS O 1s core-level spectra of the carbon-copper modulated TiO$_2$-hosts in the *bulk* and *thin-film* morphologies and TiO$_x$ (x = 1.5) XPS external standard.

The recorded XPS O 1s core-level spectra of the samples under study and $TiO_x$ (x = 1.5) reference are shown in Fig. 4. The actual BE-locations for these spectra were established as the following: $TiO_x$ = 530.4 eV, *bulk* Cx-TiO$_2$:Cu = 530.1eV and *thin-film* Cx-TiO$_2$:Cu = 530.3 eV. In order to analyze precisely the O 1s core-level line-shapes deviations from each other, we had shifted them in BE-scale employing the manner of the best match with $TiO_x$ (x = 1.5) XPS O 1s external standard, as it had been suggested for the first time as a methodical approach in XPS analyzing by J. Kawai *et. al.* [53]. The use of TiO$_2$ O1s XPS external standard for comparing seems unreasonable, because carbon-embedding before copper implantation stage had already produced some oxygen defects in the microstructure of matrices (ballistic effects of an ion-implantation), and, thus, it is more correctly to speak about Cu-implantation into Cx-TiO$_x$ matrices, rather than into Cx-TiO$_2$. As an additional valid argument for our XPS external standard selection the 529-530 eV BE-range for stoichiometric metal oxides [43-44] might be considered, whereas our XPS O 1s data is not matching it. Concretely, the stoichiometric TiO$_2$ XPS external standard gives the value of 529.8 eV for O 1s core-level, which is, from the very beginning, actually away from the neighboring BE-positions for the $TiO_x$ (x = 1.5) and Cx-TiO$_2$:Cu samples in both used morphologies so we haven't displayed it for the Figure 4 clarity.

From Figure 4 one can see the dissimilar behavior of known oxygen defects-linked 532 eV XPS sub-band. Recall, that in our case there is no background to interpret this sub-band on the basis of OH-groups, because no hydrophobic treatment of the surface of our samples which usually causes the appearance of "alien" OH$^-$ groups was used (see samples synthesis descriptions in "Samples preparation and experimental details" section). On the oxygen defects-linked basis, the *bulk* Cx-TiO$_2$:Cu has the strongest deviations from initial O – Ti bonding in untreated TiO$_2$ and even from oxygen-deficit $TiO_x$ (x = 1.5), whereas *thin-film* Cx-TiO$_2$:Cu has the smallest one (compare blue and red O 1s spectra at Fig. 4). This established strong XPS dissimilarity of O 1s core-level spectra for the *bulk* and *thin-film* morphologies well agrees with the above reported XPS Cu 2p$_{3/2–1/2}$ core-level analysis, according to which the final formal formulas are accounted as *bulk* Cx-TiO$_2$: [Cu$^+$][Cu$^{2+}$] and

*thin-film* Cx-TiO$_2$:[Cu$^+$][Cu$^0$] and that's why. The presence both of Cu$^+$ and Cu$^{2+}$ in the *bulk* Cx-TiO$_2$ means that all embedded copper ions "captured" the oxygen from TiO$_x$ oxygen-titanium lattice, increasing the oxygen deficiency in already oxygen deficient Cx-TiO$_2$:Cu up to the highest possible level (and, thus, the 532 eV sub-band has the most profiled XPS parameters in O 1s core-level). Conversely, in the *thin-film* only Cu$^+$ states are linked with oxygen, because Cu$^0$ evidently can't be bonded with it at all, so the 532 eV defective sub-band has the lowest XPS profiling (see Fig. 4). With that, the overall line-shape of XPS O 1s spectra exhibits the TiOx–character majority (i.e. oxygen–deficient character) and is quite dissimilar with that for stoichiometric copper oxides. Low concentrations of embedded copper are responsible for it (see the data in Table I).

On the first look the appearance of Cu$^0$ means Cu-losses from the ion-beam stimulated synthesis and modulation of Cx-TiO$_2$ microstructure (this effect was established earlier by the other researchers even for amorphous hosts [59,60] and in our previous findings [61,62]), but, what might be really interesting in our humble opinion, it is accompanied by embedded metal treating of oxygen vacancies (see, for instance, Ref.[63,64]), and this occurred one more time with the current samples even under sequential and relatively soft (regarding applied mode) ion-beam stimulated synthesis. Thus one might suppose about more deep and strong coupling among these two effects, because they were detected and fixed in appearing for a relatively wide set of matrices and implants of a dissimilar chemical origin, but, at the same, are pooled by the common type of nonequilibrium ion-embedding conversion and functionalization of the final microstructure of a material. Lastly, we have to note as well that it was not possible to record the XPS O 1s spectra of Cx-TiO$_2$ hosts in both morphologies before Cu-implantation step due to technical limitations of the sequential carbon-copper loadings (uninterruptable idle-cycle in vacuum before Cu-implantation without a contact with air ambient in order to prevent possible accidental additional and undesirable oxidations or/and acidifications of Cx-TiO$_2$).

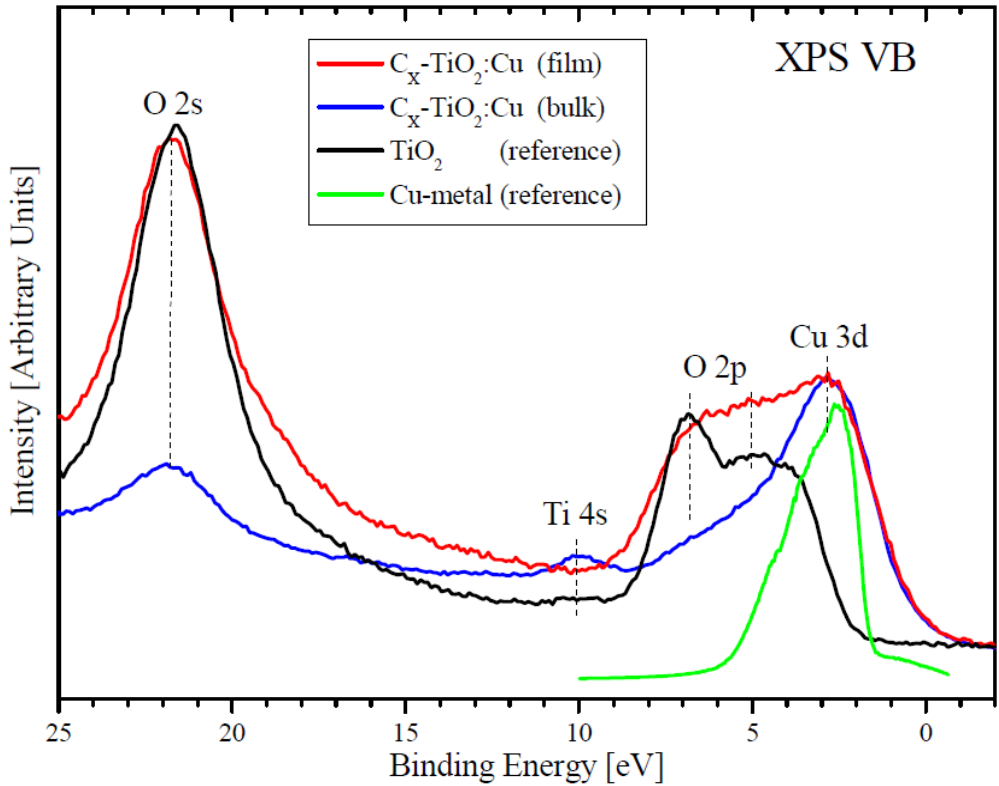

**Figure 5.** XPS Valence Band (VB) spectra of the sequentially carbon-copper modulated TiO$_2$-hosts in the *bulk* and *thin-film* morphologies, TiO$_2$ and Cu-metal [65-66] XPS external standards.

Valence band XPS spectra of the samples under study and XPS external standard references are shown in Fig. 5. On the whole, there is no at all any BE-shift of O 2s core-like (or semi-core) partial electronic states and the formal differences in O 2s band-tails are rather linked with dissimilar XPS BackGround (BG) contributions because of the dissimilar morphologies [65] than with notable contributions from ion-beam fabricated copper-oxygen clusters due to the relative low concentration of copper in compare with that for carbon and other partial components (Table I). Also we have to note one more time that according to reported above core-level analysis (please see discussion above) not all the copper have interacted with the lattice oxygen, partially remaining in a metallic form, and as a distinct visually XPS spectral signature one can see the lower O 2s signal in the VB of Cx-TiO$_2$:[Cu$^+$][Cu$^0$] *thin-film* towards the *bulk* Cx-TiO$_2$: [Cu$^+$][Cu$^{2+}$]. At the same time a little forward, exactly this contingently "the lack of copper-oxygen interactions" *thin-film* sample exhibiting a little-bit lower intensities and band-shape transformations in the O 2p region of XPS VB base-band area (see Fig. 5, the BE-range from 7 to 5 eV). Such a mentioned behavior well coincides with the offered interpretation and above reported core-levels analysis.

The BE-vicinity of 10 eV belongs to Ti 4s partial densities of states [44], and they are well-resolved by XPS in the untreated $TiO_2$-host employed as XPS external standard but not in $TiO_2$:[$Cu^+$][$Cu^0$] *thin-film* and *bulk* $Cx-TiO_2$:[$Cu^+$][$Cu^{2+}$] (Fig. 5). From Atomic Calculations of Photoionization Cross-Sections and Asymmetry Parameters [67] follows that the relation among Ti 4s, O 2p, and Cu 3d cross-sections σ for Al *K*α X-ray excitation is $0.5 \cdot 10^{-3}/0.24 \cdot 10^{-3}/1.2$, respectively, so it is not a surprise that the Ti 4s states are not resolved in XPS VB spectra of carbon-copper modulated samples. Taking into account Ref. [58] and σ (Ti 3d) = $0.17 \cdot 10^{-3}$ with a usual location at 2.2 eV and the relatively weak contribution of Ti 4s even into VB of $TiO_2$, then the suppression of Ti states in the valence band becomes evident. The VB Base-band Width (BBW) is enlarged greatly from 4.3 eV for untreated $TiO_2$ up to 7 eV because of the appeared Cu 3d contribution. With that, the BBW becomes of Cu 3d majority character with all further consequences for the material electronic properties. Earlier the dominating character of Cu 3d electronic states in the BBW region was established independently employing X-ray emission overlay approach for the triple-component oxide [68], thus proving the suppressing character of copper 3d stated appearance in BBW. No essential influence of carbon on the electronic structure of final carbon-copper modulated samples were found except C 1s core-level transformations.

**Table II.** Formation energies (in eV per defect) for various combinations of defects – oxygen vacancies (vO), substitutional carbon impurities ($C_s$), substitutional (S) and interstitial (I) copper ions – in different $TiO_2$ matrices. The most probable configurations of defects in realistic matrices (see discussion in the text) are marked with bold font.

| Host/Impurity | rutile (bulk) | anatase (bulk) | anatase (surface) |
|---|---|---|---|
| C | +8.77 | +8.58 | +3.18 in bulk<br>**-0.26** on surface |
| vO | +4.43 | +4.57 | +3.31 |
| vO after $C_S$ | **+3.95** | **+3.89** | **+2.90** |
| 1S | +6.54 | +6.06 | +4.14 |
| $C_S$ | +5.10 | +5.81 | +6.13 |
| vO | +4.41 | +4.67 | **+0.27** |
| $C_S$ + vO | +4.28 | +4.22 | +0.15 |
| 2S | +6.40 | +5,84 | +4.29 (for 3S +4.45) |
| $C_S$ | +6.15 | +4.56 | +3.90 |
| vO | +4.48 | +4.11 | +1.92 |
| $C_S$ + vO | +4.20 | +3.97 | +1.85 |
| S+I | +2.54 | +2.33 | +2.08 |
| $C_S$ | **+2.10** | **+1.74** | +1.98 |
| vO | **+1.51** | **+1.46** | **+0.41** |
| $C_S$ + vO | +1.39 | +1.36 | +0.97 |
| 2S+I | +3.27 | +3.13 | +2.82 |
| $C_S$ | +2.95 | +2.84 | +2.60 |
| vO | **+1.63** | **+1.55** | +2.25 |
| $C_S$ + vO | +1.55 | +1.49 | +1.98 |

In order to understand and discuss the results of XPS measurements on a microscopic background, we performed the set of model calculations. First of all we examined the positions of carbon impurities and their role in the formation of oxygen vacancies. Results of our calculations, presented in Table II, demonstrate the large magnitude of the formation energy for the substitutional carbon impurity both in rutile and anatase *bulk* $TiO_2$. In the vicinity of the surface, the energy required for incorporation of the carbon impurity is decreased but still remains valuable. In the case of placement of carbon atom on the surface as ad-atom (Fig. 6a) the formation energy turns to negative. Thus we can conclude that implanted carbon atoms should migrate towards the surface, being driven by the large difference in the formation energies for the *bulk* and for the vicinity of surface, with the onward yield out of $TiO_2$ grains and posterior formation of separate carbon phase. This result is in a good agreement with performed XPS structure mapping which as well evidences about the formation of graphite-like phase (Fig.2).

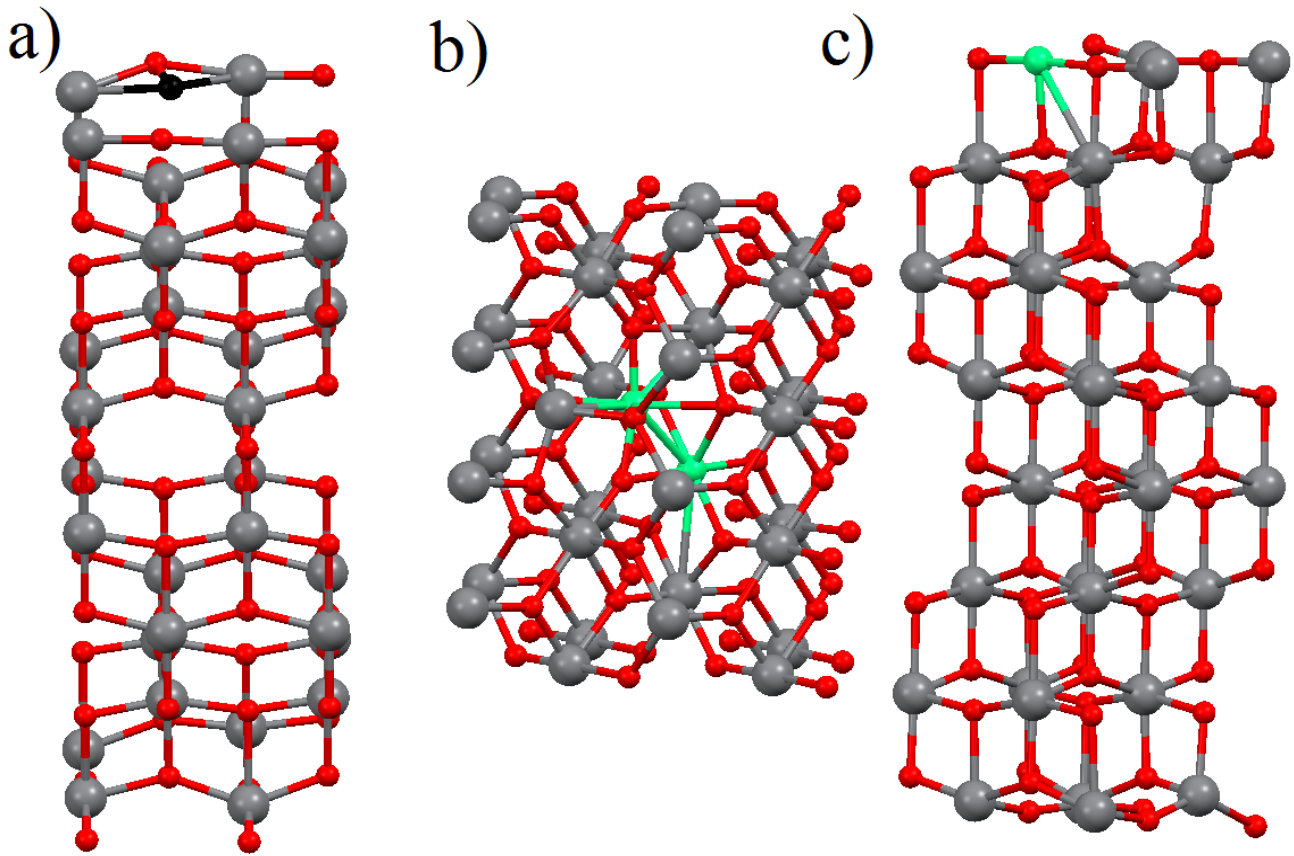

**Figure 6.** Optimized atomic structures of carbon ad-atoms: (a) on the surface (*thin-film*) of anatase $TiO_2$; (b) substitutional and interstitial copper impurities (S+I) in the *bulk* rutile $TiO_2$; (c) the combination of single substitutional copper impurity (1S) and oxygen vacancy on the surface of anatase $TiO_2$.

The formation energies of the oxygen vacancies both in rutile and anatase phases of $TiO_2$ are smaller than the energy cost of carbon implantation, but remain relatively high (see Table II). The energy difference between the defects in the *bulk* and the surface (*thin-film*) is much smaller than for the case of carbon impurities in the same positions. Calculations of the various configurations of the paired carbon impurity plus oxygen vacancy demonstrate that the lowest total energy of the system is corresponding to the closest distance between them. Therefore carbon impurities can be discussed as the centers of the formations of all configurations for copper impurities in $TiO_2$, i.e. new phase precursors. The performed calculation of the formation energies demonstrates that the presence of substitutional carbon impurity provides visible decreasing of formation energy of oxygen vacancy. Since the difference between formation energies of oxygen vacancies in the *bulk* and on the surface (*thin-film*) is not very large, so the migration of vacancies from bulk to the surface is much less favorable than for carbon impurities. This result is also in agreement with experimentally obtained XPS data, well evidencing the abundance of oxygen vacancies both in *bulk*-rutile and *thin films*-anatase after sequential carbon-copper loadings (Fig. 4).

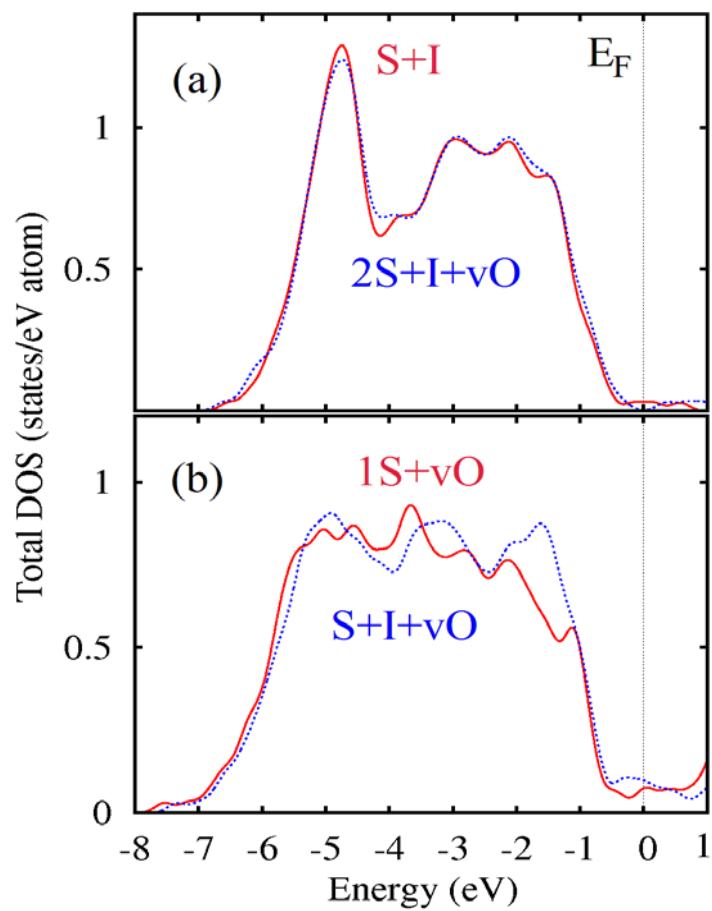

**Figure 7.** Base-band width Densities of States (DOSes) for the most energetically favorable configuration of copper defects and oxygen vacancies (a) in the *bulk* and (b) *thin-films* of $TiO_2$.

The next step of our modeling is the analysis of energetics of various copper defects in $TiO_2$ matrices with and without defects. The results of calculations (see Table II) demonstrate the insignificant difference between rutile and anatase in the *bulk* morphology. For both phases of $TiO_2$ the preferable configurations of copper impurities in the absence of other defects are the combinations of substitutional and interstitial impurities (Fig. 6b). The nature of this phenomenon is the combination of larger ionic radii and lower valences of copper that titanium that makes simple substitution energetically much less favorable. The presence of other defects provides decreasing of the formation energies for all configurations of copper defects without varying the tendency to fabricate the clusters of substitutional and interstitial copper impurities. The Base-band area electronic structure shown in Fig. 7a demonstrates that the formation of discussed clusters provides clearly visible shift of the valence bands toward Fermi level. The presence of oxygen vacancies provides at the same time insignificant changes in electronic structure thow. So the results of calculations are in agreement with XPS VB structure mapping which demonstrates the presence of metallic copper in the *bulk* rutile $TiO_2$ host-matrix.

In the case of a surface without defects, the results of our calculations demonstrate the same tendency for the cluster formation (see Table II) that are the combination of substitutional and interstitial defects, as it have been established in the *bulk* $TiO_2$. The presence of carbon impurities does not change valuable energetics of the copper defects configurations, in contrast with oxygen vacancies which create some hollow-space and make the formation of the single substitutional copper defect (Fig. 6a) also rather possible. The Base-band electronic structure (Fig. 7b) of the most probable defects demonstrates the smearing of the valence band in the *thin-film* Cx-$TiO_2$:Cu. The nature of this smearing is the presence of various types of oxygen atoms (surface, vicinity sub-surface, the environmental atoms of copper impurities, etc.) that provides the broadening of O 2p band (see Fig. 5 and 7a). At the same time the formation of copper clusters provides the regular increasing of the closest to Fermi level DOS'es. This result is also in qualitative agreement with experimental results

(Figs. 3 and 5) that demonstrate a combination of copper-oxide and metallic copper on the surface of oxygen-deficit $TiO_2$.

## 6. Conclusions

The heavy carbon-loadings with posterior Cu-sensitization of the *bulk* and *thin-film* $TiO_2$ hosts were made using sequential pulsed ion-implantation technique. The complete XPS qualification of the electronic structure of final Cx-$TiO_2$:Cu samples allows to establish the actual formulas as Cx-$TiO_2$:[$Cu^+$][$Cu^{2+}$] for the *bulk* and Cx-$TiO_2$:[$Cu^+$][$Cu^0$] for *thin-film* morphologies, respectively. No copper-carbon and titanium-carbon interactions were experimentally found due to the remaining majority of neutral C–C bonds ($sp^3$-type) after carbon-copper implantation and only a lack of embedded carbon is fabricating the O–C=O clusters. This conclusion well agrees with XPS VB BBW analysis where it have been established the absence of BBW-broadening toward Fermi level. With that, there are no doubts about the dominating majority of Cu 3d states in comparance with other XPS BBW components, especially in the valence base-band of the *bulk* Cx-$TiO_2$:Cu. The performed theoretical model-scenarios for several trends of over the threshold carbon-loadings and posterior Cu-sensitization well agrees with experimentally mapped electronic structures of final Cx-$TiO_2$:Cu samples. It was shown that neutral carbon low-dimensional impurities play the role of precursors for the new phases while Cu-sensitizing the Cx-$TiO_2$ intermediate-state hosts.

**Acknowledgements**

Ion-implantation stimulated synthesis of the samples was made under support of the Act 211 of the Government of Russian Federation (Agreement No. 02.A03.21.0006) and the Government Assignment of Russian Ministry of Education and Science(Contract No. 3.1485.2017/ПЧ). XPS measurements were made under support of Russian Science Foundation (Project 14-22-00004). DWB acknowledge support from the Ministry of Education and Science of the Russian Federation, Project №3.7372.2017/БЧ.